\documentclass{article}
\usepackage{spconf,amsmath,graphicx}
\usepackage{paralist,bbding,pifont}
\usepackage{multirow}
\usepackage{colortbl}
\usepackage{float}

\title{An adaptive filter bank based neural network approach for time delay estimation and speech enhancement}
\name{Lu Ma\thanks{Corresponding author is Lu Ma, Email: iamroad@163.com.}}
\address{}
\begin{document}
%
\maketitle
\begin{abstract}
Time delay estimation (TDE) plays a key role in acoustic echo cancellation (AEC) using adaptive filter method. Considerable residual echo will be left if estimation error arises. Here, in this paper, we proposed an adaptive filter bank based neural network approach where the delay is estimated by a bank of adaptive filters with overlapped time scope, and all the energy of filter weights are concatenated and feed to a classification network. The index with maximal probability is chosen as the estimated delay. Based on this TDE, an AEC scheme is designed using a neural network for residual echo and noise suppression, and the optimally-modified log-spectral amplitude (OMLSA) algorithm is adopted to make it robust. Also, a robust automatic gain control (AGC) scheme with spectrum smoothing method is designed to amplify speech segments. Performance evaluations reveal that higher performance can be achieved for our scheme.  
\end{abstract}
\begin{keywords}
acoustic echo cancellation, time delay estimation, gain control, adaptive filter, neural network
\end{keywords}
\section{Introduction}
\label{sec:intro}

\begin{figure*}
\centering
\includegraphics[scale=0.58]{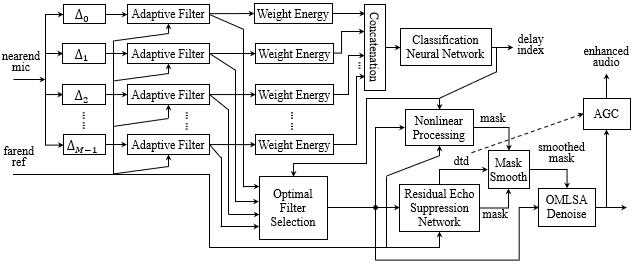}
\caption{Algorithm structure of the proposed time delay estimation and speech enhancement.} \label{fig1}
\end{figure*}

Acoustic echo arises when the microphone at the near-end picks up the loudspeaker’s sound plus its reverberation and heard by the speaker itself at the far-end. Since there is a reference signal named the far-end, adaptive filter (AF) is always employed for acoustic echo cancellation (AEC)~\cite{ref5,ref6}., i.e., the echo is estimated by adaptive filter based on the far-end signal and then subtracted from the near-end mixture. However, nonlinear distortion may be introduced to acoustic echo in practice, leading to considerable residual echo. To handle this problem, some methods have been proposed, such as Nonlinear AEC (NLAEC) method by using a set of nonlinear basis functions for echo estimation~\cite{ref7,ref8} and nonlinear post-filtering method by cascading an additional nonlinear signal processing module for residual echo suppression~\cite{ref12,ref13}. Recently, since its great potential in speech processing tasks, neural network (NN) has been used for echo cancellation or suppression, such as NN-based postfiltering method~\cite{ref14,ref15,ref16,ref18}, NN-based NLAEC method~\cite{ref19}, separation-based method~\cite{ref20}, end-to-end approach~\cite{ref17,ref21}. 

Nevertheless, as is revealed in~\cite{ref14,ref15,ref16}, a joint approach of linear filtering and neural network based RES has shown to be more effective. Here, in this paper, we adopted this paradigm while pay our attention to the time delay estimation (TDE) due to its key role in adaptive filter based AEC. If the estimated time differs from the actual one, the adaptive filter should process such error with additional filter length, in turn reducing the filter length that can be left for actual echo estimation. Generally, cross-correlation function (CCF) is used for TDE as is done in WebRTC AEC\footnote{{https://github.com/HackWebRTC/webrtc}}. The accuracy could be vulnerable if nonlinear echo and complex noise are introduced. To overcome this problem, we proposed an adaptive filter bank based neural network for TDE. A bank of adaptive filters with overlapped time scope between adjacent filters are employed for linear echo cancellation. The weight energy of each filter is calculated, and all these weight energy are concatenated together and feed to a pre-trained classification neural network for time delay estimation. The index with the maximal probability is considered as the delay between the near-end and the far-end.

Base on this TDE, an AEC scheme is designed using neural network for residual echo and noise suppression. To make the suppression more robust that giving as less distortion as possible to the near-end speech, the results of the suppression network is not directly performed on the near-end spectrum. Instead, it is first be smoothed with the gain of nonlinear signal processing (NLP) where the NLP can be referred to WebRTC AEC, and then this smoothed gain is used to guide the optimally-modified log-spectral amplitude (OMLSA) algorithm~\cite{ref23} for noise estimation and suppression by considering the residual echo as a special noise. Also, an automatic gain control (AGC) scheme where gain smoothing method is designed for mitigating spectrum discontinuity that would be introduced by AGC. Simulation results reveal the effectiveness and superiority of our proposed schemes. It is notable that all the algorithms mentioned in our proposed method can be realized based on open-sourced codes\footnote{{https://github.com/xiph/speex}}\footnote{{https://github.com/xiph/rnnoise}}\footnote{{https://github.com/xiaochunxin/OMLSA-MCRA}}, making it convenient for work reproduction and implementation. 

\section{Framework}
\label{sec:format}
The framework is illustrated in Fig.~\ref{fig1}. Three modules are included, i.e., time delay estimation, residual echo suppression and automatic gain control. The time delay is first estimated by the adaptive filter banks with classification neural network, then the optimal filter output is selected and used for residual echo suppression with neural network and denoising with OMLSA. AGC is then used to amplify the audio. 

\subsection{Time Delay Estimation}
\label{ssec:tde}
As is is shown in Fig.~\ref{fig1}, assume that there is $K$ samples per frame, and the length per filter is $N$ frames, the overlap between adjacent filters is $L$ frames, and totally $M$ filters are used. A data buffer whose length is $M \times N - L$ denoting the maximal time delay that can be estimated is used for the far-end signal, and each time the newest far-end frame is shifted into buffer with the oldest frame shifted out. From the left to the right, with $N - L$ spacing, copy the data at the corresponding position to the corresponding filter. The total length per filter is $K \times N$, and is splitting into $N$ blocks with $K$ samples per block. For each filter, these $K$ weight energy values are sum together as the block energy, and totally $N$ block energy are obtained per filter. Therefore, totally $M \times N$ block energy are obtained for all filters and they are concatenated together for feeding to the classification network. Here, MDF~\cite{ref6} is employed, $N = 32$ blocks is used for each filter with $L = 8$ blocks overlapping between adjacent filters, and totally $M = 5$ filters are used. By feeding the $M \times N$ ($M \times N = 5×32$) block energy values to the classification network giving $M \times N - L$ categories, and each category denotes the corresponding delay. The one with the maximal probability is considered as the estimated delay. The structure of the classification neural network is extracted from the voice activity detection (VAD) path of RNNoise~\cite{ref24} whose input is the concatenated energies and training with cross-entropy loss.


\begin{figure}
\centering
\includegraphics[scale=0.26]{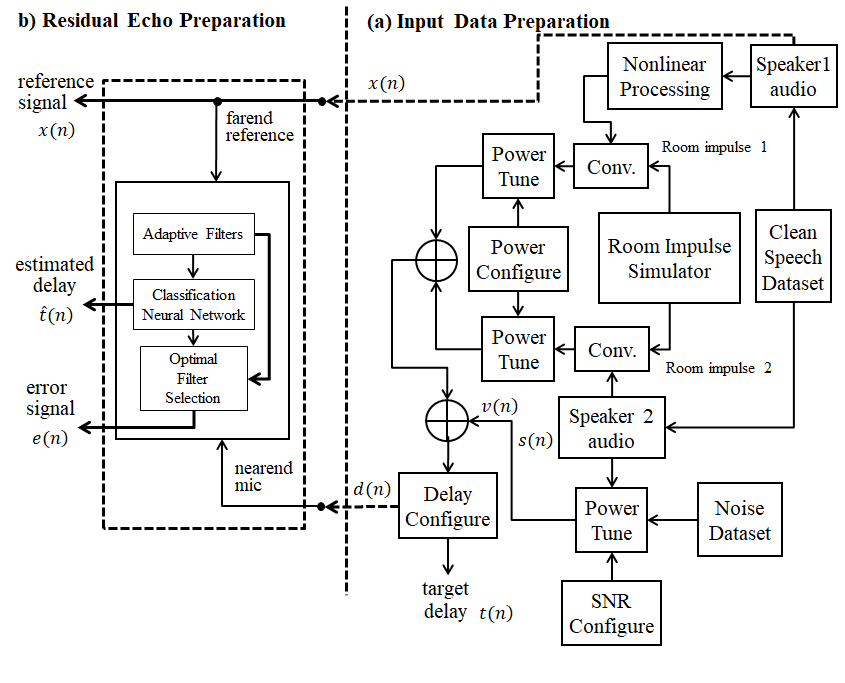}
\caption{Schematic of TDE training data preparation.} \label{fig3}
\end{figure}

The preparation of training data for delay classification network is illustrated in Fig.~\ref{fig3}(a). Firstly, two audio clips from two different speakers are randomly chosen as the far-end and the near-end audios. The far-end signal denoted by $d \left( {x} \right)$ is then processed by the nonlinear processing (NLP) module. Two room impulse response (RIR) files are randomly selected from the dataset, and then convolved with the near-end clean signal $s \left( {n} \right)$ and the nonlinear far-end signal for emulating the near-end speech and the echo signal. These two signals are then power tuned according to the predefined speech-to-echo ratio (SER) and added together for representing the mixture speech received by microphone. At the same time, a noise clip is randomly selected from noise dataset and then power tuned with respect to the near-end speech power based on the predefined signal-to-noise ratio (SNR). It is then added on the mixed speech denoted by $d \left( {n} \right)$. The noisy mixture is then delayed by a predefined value $\tau \left( {n} \right)$ for emulating the delay between the near-end microphone and the far-end reference. The generated near-end mixture $d \left( {n} \right)$ and the far-end reference $x \left( {n} \right)$ are feed to the batch of adaptive filters whose weights energy are calculated and concatenated and then inferenced by the classification network, getting an estimated delay denoted by $\hat{\tau} \left( {n} \right)$.

\subsection{Residual Echo Suppression}
\label{ssec:res}
\begin{figure}[htb]
\centering
\includegraphics[scale=0.28]{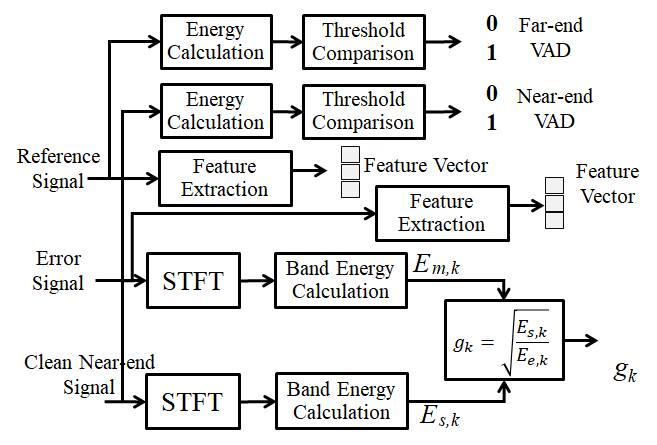}
\caption{Data preparation for suppression network.} \label{fig4}
\end{figure}

\textbf{Suppression neural network}
As is shown in Fig.~\ref{fig3}(b), the time delay is first estimated and then the optimal filter output is considered as the error signal $e$$\left({n}\right)$. Therefore, the neared clean signal $s$$\left({n}\right)$, the far-end signal $x$$\left({n}\right)$ and the error signal $e$$\left({n}\right)$ are prepared as is shown in Fig.~\ref{fig4}. The near-end clean signal and the error signal are first converted to frequency domain by short-time Fourier transformation (STFT), and then split into frequency bands based on mel-cepstrum scale, the gain per band between the near-end signal and the error signal is computed as $g_k = \sqrt{E_{s,k}/E_{e,k} } $ (where $E_{s,k}$ and $E_{e,k}$ are the energy of the $k$-th band of the near-end and the error signal respectively). This is the target band gains that are used for training. At the same time, the frame energies of the far-end and the near-end are calculated and compared to predefined threshold for voice activity detection (VAD) (i.e., `1' represents speech exists and vice versa). These are used as target of double talk detection (DTD). The 39 dimensions MFCC features (i.e., 13 dimensions MFCC and its first and second order difference) of the far-end and the error signal are extracted and concatenated as input to RNNoise~\cite{ref24} where the VAD is replaced by DTD in our scheme. 

\textbf{Denoising with OMLSA} The residual echo is considered as a special noise and would be suppressed by OMLSA algorithm. Te power spectrum of the input error signal $e$$\left({n}\right)$ is calculated and feed to the noise estimation as is shown in Fig. 1 of~\cite{ref23}. The band gains estimated by the suppression neural network are first interpolated to obtain frequency gains (named mask in the figure) and then smoothed with that of NLP. The smoothed gain is compared with threshold to decide whether this frequency is noise or not, guiding to estimate noise.  Specifically, the noise decision flag $I\left({k,l}\right)$ of the $k$-th frequency at the $l$-th frame can be identified by, 
\begin{equation}
I(k, l)=\left\{\begin{array}{l}
1, \text { if } G(k, l) \geq 0.5 \\
0 \text {, if } G(k, l)<0.5
\end{array}\right.
\end{equation}
Therefore, the speech conditional present probability $\hat{p}^{'}\left({k,l}\right)$ can be obtained by smoothing with factor $\alpha_{p} \left({0 \leq \alpha_{p} \leq 1}\right) $ as
\begin{equation}
\hat{p}^{\prime}(k, l)=\alpha_p \cdot \hat{p}^{\prime}(k, l-1)+\left(1-\alpha_p\right) \cdot I(k, l)
\end{equation} 
After some deductions as is referred to in~\cite{ref23}, the OMLSA frequency gain $G_{\text {o }}(k, l)$ is obtained by Eq. (16) in~\cite{ref23} and then it is smoothed by the neural network gain as,
\begin{equation}
G_{\text {f}}(k, l)={p}(l,k) \cdot G_{\text{o}}(k, l)+\left(1-{p}(l,k)\right) \cdot G_{\text {n }}(k, l)\\
\end{equation}
where $G_{\text{o}}\left({k,l}\right)$, $G_{\text{n}}\left({k,l}\right)$ and $G_{\text{f}}\left({k,l}\right)$ are the frequency gain of the OMLSA, the suppression network and the smoothed one respectively, ${p}\left({k,l}\right) = \hat{p}(l,k)$ if the current frame is speech, otherwise ${p}\left({k,l}\right) = \hat{p}(l)$ is the near-end speech present probability obtained by RNNoise.

\subsection{Automatic Gain Control}
\label{ssec:agc}
For each frame, the maximal amplitude $\hat{A}_k$ and the mean amplitude $\bar{A}_k$ is calculated. Then, the corresponding gains are calculated as $\bar{G}_k={A_{\text {mean }}{\text {def }}}/{\bar{A}_k}, \hat{G}_k={A_{\max }^{\text {def }}}/{\hat{A}_k}$ based on the predefined mean $A_{mean}^{def}$ and maximum $A_{max}^{def}$. The minimal gain between these two values is obtained as the current frame gain, i.e., $G_k=\min \left(\bar{G}_k, \hat{G}_k\right)$. This gain value is smoothed frame by frame with $\alpha \left({0 \leq \alpha \leq 1}\right) $ by, $G_k=\alpha \cdot G_{k-1}+(1-\alpha) \cdot G_k$. This smoothed gain is then restricted by the maximal gain for saturation guarding as $G_k=\min \left(G_k, \widehat{G}_k\right)$. Unfortunately, the gains for adjacent two frames would be discontinuous due to saturation guarding, leading to discontinuous spectrum. To deal with this problem, a gain discontinuous control scheme by smoothing with sigmoid function is proposed and formulated by,
\begin{equation}
\begin{aligned}
G_k(m)=G_{k-1}+\frac{\Delta G_k}{1+e^{-x}}, \Delta G_k=G_k-G_{k-1}\\
x=-5.0+\frac{10.0}{M} * m, m=0,1,2, \ldots, M-1\\
\end{aligned}
\end{equation}
where $G_{k-1}$ and $G_k$ are gains of two adjacent frames, $M$ is the number of samples that is being smoothed, $ e^{-x}$ is exponential function with $x$ as the argument, $G_k(m)$ denotes the gain of the $m$-th sample at the $k$-th frame.

\begin{figure}[htb]
\centering
\includegraphics[scale=0.4]{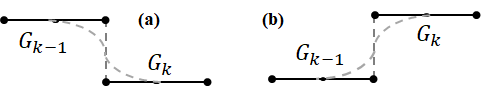}
\caption{Depiction of gain discontinuity.} \label{fig8}
\end{figure}

As is depicted in Fig.~\ref{fig8}, two kinds of gain discontinuous are considered: 1) The gain of current frame is smaller than the precious one as is marker by the dotted curve in Fig.~\ref{fig8}(a), smoothing the last $M$ samples of the previous frame from $G_{k-1}$ to $G_k$ using sigmoid function. b) The gain of current frame is greater than the precious one as is marker by the dotted curve in Fig.~\ref{fig8}(b), smoothing the  first $M$ samples of the current frame from $G_{k-1}$ to $G_k$ using sigmoid function.

\section{Experiments}
\label{sec:exp}
\subsection{Dataset Preparation}
\label{ssec:dataset}
The dataset of AEC challenge 2021 is utilized for AEC model training. The preparation is referred to in~\cite{ref21}. Only the far-end and the echo signals are used and cut into chunks of 4 s. Clean speech and noise corpus from DNS-Challenge 2021~\cite{ref27} are used. The near-end signal is convolved with RIR randomly selected from~\cite{ref28}. For each impulse response, the begin of the direct path was identified and set to position 0 as proposed in~\cite{ref29}. The reverberant near-end was mixed with the echo signal with SER randomly selected from -30 dB to 30 dB with 5 dB spacing. This mixture is then added by noise signal with SNR randomly selected from -10 dB to 30 dB with 5 dB spacing. Since delay is unknown in advance for the AEC challenge 2021 dataset, we synthesized the TDE dataset as is illustrated in Fig.~\ref{fig4}(a). The nonlinear model referred to in~\cite{ref20} is used as the nonlinear processing in Fig.~\ref{fig3}(a). The time delay is randomly controlled from 0 to 500 ms with 10 ms spacing. In this way, $2\times10^5$ clips are generated for TDE training. These files are also used for AEC training, i.e., totally $4 \times 10^5$ files are used for AEC training. Splitting into subsets with percent of 7:2:1 for train, valid and test.

\subsection{Results Analysis}
\label{ssec:results}
\textbf{\emph{TDE}} Three TDE schemes were compared here as is listed in Table~\ref{tab1}, i.e., WebRTC AEC, WebRTC AEC3 and our proposed scheme (named ADF-NN-TDE here). The number inside the backet is the tolerance of delay estimation error. As can be seen from the table that, our proposed scheme can get the best performance. Moreover, with the tolerance of estimation error decreases, our scheme decreases a little, while that of the WebRTC AEC scheme drops a lot. By comparing with that of the WebRTC AEC3, it reveals that high robust is possessed in our scheme by replacing the complex manual designed postprocessing with a lightweight neural network. 

\textbf{\emph{AEC}} Here, we compared our scheme with two classical methods, i.e., the WebRTC AEC (referred to as `WebRTC') and the end to end neural model proposed in~\cite{ref21} (referred to as `End2End'), `TDE+NLP' refers to our proposed method only with NLP for residual echo suppression, `++NN' refers to the proposed method where RES network is added on `TDE+NLP', `+++OMLSA' refers to the method where OLMSA is also added. As can be seen from Table~\ref{tab2}, by introducing neural network to TDE, the audio quality can be increased a lot. Though the WebRTC scheme can get minimal variance, the mean PESQ gain is small. Moreover, by introducing neural network for RES and smoothing the gain with that of the NLP using the near-end speech probability, the AEC performance can be improved further as is figured in `++NN' scheme. Specifically, employing OMLSA as post-processing, the spectrum distortion can be further reduced with the mean value of PESQ gain improved and the variance value decreased. Fortunately, the overall neural model can be controlled in a considerable size. It also can be seen from Table~\ref{tab2} that, though the `End2End' scheme gain the minimal variance and considerable mean value, its model size is 2.3 times larger than our proposed schemes. 

\begin{table}
\centering
\caption{Comparisons of different TDE scheme.}\label{tab1}
\begin{tabular}{|c|c|c|}
\hline
Schemes &  Acc. ($\pm 25$ms) & Acc. ($\pm 5 $ms)\\
\hline
WebRTC AEC &  {73.04$\%$} & {50.34$\%$}\\
WebRTC AEC3 &  {80.32$\%$} & {72.56$\%$}\\
ADF-NN-TDE & {91.67$\%$} & {89.88$\%$}\\
\hline
\end{tabular}
\end{table}

\begin{table}
\caption{AEC performance comparison.}\label{tab2}
\centering
\arrayrulecolor{black}
\begin{tabular}{!{\color{black}\vrule}l!{\color{black}\vrule}l!{\color{black}\vrule}l!{\color{black}\vrule}l!{\color{black}\vrule}l!{\color{black}\vrule}l!{\color{black}\vrule}} 
\hline
\multirow{2}{*}{Schemes} & \multicolumn{4}{c!{\color{black}\vrule}}{PESQ Gain} & \multirow{2}{*}{\begin{tabular}[c]{@{}c@{}}Model\\Size\end{tabular}}  \\ 
\cline{2-5}
 & Mean   & Min.    & Max.   & Var.  &   \\ 
\hline
End2End  & \textbf{0.545} & 0.002  & 1.508 & \textbf{0.068}    & \textbf{1827K}   \\ 
\hline
WebRTC    & 0.138\textbf{} & -0.275 & 0.876 & 0.038\textbf{}    & ~  \\ 
\hline
TDE+NLP   & 0.472  & -0.148 & 1.579 & 0.205  & 468K  \\ 
\hline
++NN   & 0.572  & -0.112 & 1.879 & 0.162  & 794K  \\ 
\hline
+++OMLSA  & \textbf{0.633} & -0.103 & 1.732 & \textbf{0.101}   & \textbf{794K}  \\
\hline
\end{tabular}
\arrayrulecolor{black}
\end{table}

\begin{figure}[htb]
\centering
\includegraphics[scale=0.46]{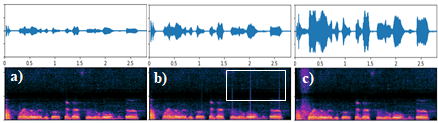}
\caption{Schematic of gain smoothing.} \label{fig10}
\end{figure}

\textbf{\emph{AGC}} As is shown, Fig.~\ref{fig10}(a) is the original wave, Fig.~\ref{fig10}(b) and Fig.~\ref{fig10}(c) are the one without and with gain smoothing. Obviously, without smoothing, discontinuity will be introduced to the spectrum as is marked in Fig.~\ref{fig10}(b) with rectangular box. This discontinuity would be almost arise when the amplitude of adjacent frames differ much, especially when the previous frame amplitude is much smaller than that of the current frame. In consequence, the gain cannot be enlarged too much for accommodating such discontinuity. Contrarily, the spectrum in Fig.~\ref{fig10}(c) is consecutive by using gain smoothing and the gain can be increased rapidly to a large scale.

\section{Conclusion}
Inspired by WebRTC AEC3, a time delay estimation scheme based on a bank of adaptive filters and a lightweight classification neural network was proposed, resulting in higher accuracy and more robust in term of time delay estimation. This TDE was then employed for AEC together with a suppression neural network for residual echo and noise suppression, and OMLSA algorithm was adopted to obtain superior enhancement performance. Moreover, by smoothing the amplitude gain with sigmoid function, a robust AGC was obtained.

\vfill\pagebreak

\bibliographystyle{IEEEbib}
\bibliography{ref1,ref2,ref3,ref4,ref5,ref6,ref7,ref8,ref9,ref10,ref11,ref12,ref13,ref14,ref15,ref16,ref17,ref18,ref19,ref20,ref21,ref22,ref23,ref24,ref25,ref26,ref27,ref28,ref29}

\end{document}